\documentclass[
    aps,
    prd,            
    reprint,        
    superscriptaddress,
    nofootinbib,
    amsmath,
    amssymb,
    aps
]{revtex4-2}

\usepackage[T1]{fontenc}
\usepackage{graphicx}
\usepackage{bm}
\usepackage{physics}
\usepackage{hyperref}
\usepackage{dcolumn}
\usepackage{booktabs}
\usepackage{multirow}
\usepackage{slashed}
\usepackage[dvipsnames]{xcolor}
\usepackage[caption=false]{subfig}
\usepackage{float}
\usepackage[normalem]{ulem}

\begin{document}


\title{
Reconstruction of Dynamical Dark Energy and Gravity from latest cosmological data
}

\author{Maria Luisa Gallavotti}
\email{gallavotti@lorentz.leidenuniv.nl}
\affiliation{Instituut-Lorentz, Leiden University, PO Box 9506, Leiden 2300 RA, The Netherlands}

\author{Alessandra Silvestri}
\email{silvestri@lorentz.leidenuniv.nl}
\affiliation{Instituut-Lorentz, Leiden University, PO Box 9506, Leiden 2300 RA, The Netherlands}

\date{\today}


\begin{abstract}
Recent cosmological observations have renewed interest in non-standard late-time evolution. We investigate whether departures from the standard expansion history are accompanied by modifications of gravitational dynamics through a joint, non-parametric reconstruction of the effective dark energy density, $\Omega_{\rm X}(a)$, and the gravitational response functions $\mu(a)$ and $\Sigma(a)$. Combining CMB, DESI DR2 BAO, Pantheon+, and large scale structure observations, including DES Year 3 $6\times2$pt data, we perform reconstructions with and without a theory motivated correlation prior derived from stable Horndeski models. We find a persistent maximum in $\Omega_{\rm X}$ at $z\simeq0.5$--$0.7$, corresponding to a crossing of the effective phantom divide. This feature is already present in the CMB+BAO combination and remains robust when modifications to clustering and lensing are allowed. The reconstructed gravitational response exhibits suggestive enhancements in $\mu$ and $\Sigma$ over approximately the same redshift range, although these departures are less significant. Including LSS information instead favours a reduced lensing response, $\Sigma<1$, while leaving the background feature largely unchanged. An interpretation within Generalized Brans-Dicke theories illustrates how modifications to the effective gravitational coupling and a scalar-mediated fifth force may contribute to these trends. Our results indicate that current departures from $\Lambda$CDM are most significant in the expansion history, while leaving open the possibility of correlated modifications to gravitational dynamics that future Stage IV surveys can test.
\end{abstract}

\maketitle

\section{Introduction}

The observational results from the Dark Energy Spectroscopic Instrument (DESI)~\cite{DESI:2024mwx,DESI:2025zgx} have ushered in a new era of precision cosmology and renewed interest in dynamical dark energy (DE) models.  More recently, the Dark Energy Survey (DES)~\cite{DES:2026jmi} has combined its Year 6 measurements with other cosmological probes  producing some of the most stringent cosmological constraints to date. These analyses continue to exhibit mild indications of departures from the standard $\Lambda$CDM expansion history, with a preference for dynamical DE at about $3\sigma$ when the DE equation of state is parametrized using the Chevallier--Polarski--Linder (CPL) form~\cite{DES:2026jmi}.

Phenomenological parametrizations such as CPL provide a useful and economical framework to capture departures from $\Lambda$CDM, but they necessarily restrict the space of possible histories. Non-parametric reconstructions offer a complementary approach: rather than selecting a particular time dependence a priori, they allow the data to determine which features of the expansion history are actually constrained. Recent examples in the context of DESI can be found in Refs.~\cite{DESI:2024aqx,DESI:2024mwx}. Such reconstructions are particularly useful for assessing whether an apparent preference for evolving DE is robust to parametrization choices and for localizing the redshift range from which the preference originates. 

A reconstruction of the background expansion alone, however, does not provide a complete description of the relevant late-time physics. Constraints on the gravitational response must also account for degeneracies with the expansion history and the matter density~\cite{Ruiz-Zapatero:2022xbv}. A time-dependent effective DE component can also be accompanied by non-trivial dynamics of cosmological perturbations. This becomes especially pertinent when the reconstructed expansion history corresponds to an effective equation of state evolving across the phantom divide, $w_{\rm X}=-1$. Realizing such behaviour in a theoretically consistent framework generally requires additional degrees of freedom or interactions---for example non-minimal couplings to gravity or derivative self-interactions---which can modify the growth of structure and the relation between matter and the metric potentials~\cite{Creminelli:2008wc}. Nonminimally coupled scalar field models provide concrete examples of effective phantom crossing~\cite{Wolf:2025jed}. Using non-parametric reconstructions of the background and perturbations within the landscape of Horndeski theories, it was shown that DESI results hint at a non-minimally coupled gravity~\cite{Ye:2024ywg, Ye:2026yqk}.  More recently, within scalar-tensor constructions in which the DE sound speed controls the gravitational response,  the DESI  preference for a phantom-crossing equation of state has been shown to induce correlated departures from GR in the effective gravitational coupling~\cite{Reboucas:2026ljj}. Such analyses, however, necessarily impose theory-dependent relations between the background and perturbative sectors. Here we instead ask whether analogous correlations are selected directly by the data when the expansion, clustering, and lensing histories are reconstructed simultaneously yet independently. 

We address this question by jointly reconstructing three functions of time: $\Omega_{\rm X}(a)$, which parametrizes the effective contribution responsible for late-time acceleration in the background expansion, and $\mu(a)$ and $\Sigma(a)$, which parametrize departures from the $\Lambda$CDM predictions for matter clustering and gravitational lensing, respectively. This approach allows us to assess the robustness of the reconstructed background and gravitational histories, identify the departures constrained by current observations, and investigate their possible physical interpretation.

A non-parametric reconstruction of these three sectors was previously developed and applied in Refs.~\cite{Pogosian:2021mcs,Raveri:2021dbu}. Here we revisit this framework with recent cosmological data, including DESI DR2 BAO and, for the first time in this context, the DES Year 3 galaxy-clustering and weak-lensing measurements together with their cross-correlations with CMB lensing through the full $6\times2$pt likelihood. We perform the analysis both with and without a theory-motivated correlation prior derived from stable Horndeski models. The prior is used to introduce a theoretically motivated  smoothness among neighbouring reconstruction nodes; it does not require the reconstructed histories to correspond to any particular Horndeski model. 

An important challenge in high-dimensional
non-parametric reconstructions is distinguishing
features genuinely constrained by observations from
those dominated by the prior or by poorly measured
directions. We therefore complement the reconstruction with covariant principal-component analysis (CPCA), 
quantifying the reduction in uncertainty relative to the correlation prior both within each reconstructed sector and in the combined reconstruction space. Here, we examine
the displacements and redshift structure of
the individual-sector modes, while a detailed
analysis of the combined modes and their
composition is presented in a companion
Letter~\cite{CompanionPRL}.
This provides a data-driven measure of the effective number and redshift structure of the late-time cosmological modes accessible to present observations.

Beyond the reconstructed histories themselves, we
investigate their implications for standard cosmological
parameters and examine combinations of $\mu$ and
$\Sigma$ motivated by scalar--tensor theories. In
particular, we consider the gravitational slip and
explore an interpretation within Generalized
Brans--Dicke models, where modifications to the
effective gravitational coupling and the contribution
of a scalar-mediated fifth force can be distinguished.
This provides a useful theoretical perspective on
the reconstructed gravitational response without
imposing such a model in the inference. 

Finally, we perform a controlled sequence of auxiliary reconstructions in which only $\Omega_{\rm X}(a)$ is allowed to vary while $\mu=\Sigma=1$. By progressively combining primary CMB and DESI BAO data with supernovae distances, CMB lensing, and late-time structure information, we isolate which observations generate and sharpen the reconstructed low-redshift feature in the effective DE density. This comparison directly addresses the question of whether the apparent phantom-crossing behaviour is primarily selected by the CMB+BAO determination of the background cosmology, by the supernova data, or by the addition of lensing and LSS information. 

The paper is organized as follows. In Sec.~\ref{sec:th_framework} we
introduce the phenomenological framework describing
departures from $\Lambda$CDM in the background and
linear perturbations. Section~\ref{sec:methodology} presents the
non-parametric reconstruction, the Horndeski
correlation prior, the datasets, and the statistical
tools used in the analysis. In Sec.~\ref{sec:results} we discuss
the joint reconstruction, the data-constrained modes,
the impact on standard cosmological parameters,
and the interpretation of the gravitational response
within scalar--tensor theories. Section~\ref{sec:background_recon} presents
the auxiliary background-only reconstructions,
while Sec.~\ref{sec:conclusions} summarizes our conclusions.
\section{Theoretical Framework}
\label{sec:th_framework}
We start from a spatially flat Friedmann--Lemaître--Robertson--Walker (FLRW) spacetime and consider scalar perturbations around it, working in Newtonian gauge,
\begin{equation}
ds^2=-(1+2\Psi)dt^2+a^2(t)(1-2\Phi)d\vec{x}^2,
\end{equation}
where $a(t)$ is the scale factor, while $\Psi(t,\vec{x})$ and $\Phi(t,\vec{x})$ denote the Newtonian and curvature potentials, respectively.

At the background level, the expansion history is determined by the Friedmann equation, while at linear order the metric potentials are sourced by matter perturbations. Rather than specifying a particular DE or MG model, we adopt the phenomenological description that captures possible departures from $\Lambda$CDM in both the background expansion and the evolution of scalar perturbations in terms of three functions, $\Omega_{\rm X}$, $\mu$, and $\Sigma$, defined through
\begin{eqnarray}
H^2(a) &=& H_0^2\left[
\frac{\Omega_{\rm r}}{a^4}
+\frac{\Omega_{\rm m}}{a^3}
+\Omega_{\rm X}(a)
\right]\,, \\
k^2\Psi &=&
-\frac{a^2}{2M_P^2}\,
\mu(a,k)\,\rho_m\Delta_m\,, \\
k^2(\Phi+\Psi) &=&
-\frac{a^2}{M_P^2}\,
\Sigma(a,k)\,\rho_m\Delta_m\,.
\end{eqnarray}
Here $H\equiv d\ln{a}/dt$ is the Hubble rate, $H_0$ its present-day value, and $\Omega_{\rm r}$ and $\Omega_{\rm m}$ are the present-day radiation and matter density parameters.  $\Omega_{\rm X}$ describes  any contribution to the Friedmann equation other than  matter and radiation with  a cosmological constant corresponding to $\Omega_{\rm X}(a)=\Omega_\Lambda= \mathrm{const}$.
The corresponding effective equation of state is
\begin{equation}
\label{eq:EoS}
w_{\rm X}(a)=-1-\frac{1}{3}
\frac{d\ln\Omega_{\rm X}}{d\ln a}\,.
\end{equation}
A non-monotonic evolution of $\Omega_{\rm X}$ can correspond to an effective equation of state crossing the phantom divide, $w_{\rm X}=-1$.
We emphasize that $w_{\rm X}$ is an effective equation of state defined from the reconstructed expansion history, and need not correspond to that of a separately conserved dark-energy fluid. In particular, within modified-gravity theories, $\Omega_{\rm X}$ can encode contributions arising from the non-minimal gravitational coupling.

At the level of linear perturbations, $\Delta_m$ denotes the comoving matter density perturbation, while the functions $\mu(a,k)$ and $\Sigma(a,k)$ parameterize modifications to the gravitational response. The function $\mu$ controls the relation between matter perturbations and the Newtonian potential $\Psi$, and hence directly affects the growth of structure. The function $\Sigma$, instead, determines the Weyl potential $\Phi+\Psi$, which governs the deflection of light and is therefore directly probed by weak gravitational lensing. General Relativity with a minimally coupled smooth DE component corresponds to $\mu(a,k)=\Sigma(a,k)=1$.

The triplet $\{\Omega_{\rm X},\mu,\Sigma\}$ provides a convenient, and separable, phenomenological description of the background expansion, the clustering of matter, and gravitational lensing. While in a given model these functions will generally be related, in our analysis we opt for remaining agnostic  and thus do not impose a model-dependent relation among these three sectors \footnote{With the small exception of the theory correlation prior used to regularize poorly constrained temporal structure, as we will describe in Section~\ref{sec:theory_prior}.}. This allows the data to determine independently whether departures from $\Lambda$CDM arise predominantly from the expansion history, from the dynamics of cosmological perturbations, or from correlated modifications of both.

For a given theory, analytical expressions for $\mu$ and $\Sigma$ can often be derived in the quasi-static approximation, valid on scales well inside the horizon where spatial gradients dominate over time derivatives of the additional degrees of freedom and metric potentials~\cite{Silvestri:2013ne}. Beyond this approximation, their effective form can instead be obtained from the numerical solution of the full system of linear perturbation equations, as implemented in Einstein-Boltzmann solvers. When confronting these functions with cosmological data, it is common to assume specific parametric forms for their time and scale dependence.

Here, instead, we adopt a non-parametric description of $\mu$ and $\Sigma$ and do not rely on the quasi-static approximation. The functions $\mu$ and $\Sigma$ are specified directly via the binning procedure described in the next Section and consistently incorporated into the full Einstein--Boltzmann evolution of linear perturbations, retaining their complete time dependence. In the present analysis, we restrict the modifications to be scale-independent, i.e. $\mu=\mu(a)$ and $\Sigma=\Sigma(a)$. 
\section{Methodology}
\label{sec:methodology}
In this Section, we describe the methodology used to reconstruct the late-time evolution of the background, clustering, and lensing sectors without assuming a specific functional form for $\Omega_{\rm X}$, $\mu$, or $\Sigma$. We first introduce the binning and smoothing strategy and discuss its numerical implementation. We then present the data sets and likelihoods used in the analysis, before describing the statistical tools employed to identify the combinations of reconstructed parameters that are best constrained by the data and to quantify their departures from the $\Lambda$CDM expectation.

\subsection{Non-parametric reconstruction}
\label{sec:non_par_recon}
Each of the three functions $\Omega_{\rm X}(a)$, $\mu(a)$, and $\Sigma(a)$ is represented by its values at 11 nodes in scale factor. Ten of these nodes are uniformly spaced over $a\in[0.25,1]$, corresponding to $z\in[0,3]$, and an additional node is placed at $a=0.2$ ($z=4$). We collect the node values into the reconstruction vector
\begin{equation}
    \mathbf{f}\equiv
    \{\Omega_{{\rm X},i},\mu_i,\Sigma_i\},
    \qquad i=1,\ldots,11\,,
    \label{eq:reconstruction_vector}
\end{equation}
 which contains a total of 33 node values. All eleven $\mu$ and $\Sigma$ nodes as well as the first ten $\Omega_{\rm X}$ nodes are independently sampled, while the present-day value $\Omega_{\rm X}(a=1)$ is derived from the flatness condition.

We use a cubic spline to define the continuous functions between the nodes. The node grid sets the temporal resolution of the reconstruction,  while the spline supplies a continuous and differentiable representation without statistically correlating the node amplitudes. Interpolated function values nevertheless share contributions from neighbouring nodes and are consequently correlated.
At $z>4$, we add nine anchor nodes to the reconstruction grid. These nodes are uniformly spaced in scale factor over $0\leq a<0.2$, and their values are fixed by a hyperbolic-tangent profile connecting the first reconstructed node at $a=0.2$ to the early-time $\Lambda$CDM limits, $\mu=\Sigma=1$ and $\Omega_{\rm X}=\Omega_{\rm DE}={\rm const}$. The cubic spline is then constructed over the complete grid, ensuring a smooth transition without introducing additional free parameters.

\subsection{Theory correlation prior}
\label{sec:theory_prior}
A reconstruction in a high-dimensional node basis can overfit statistical fluctuations in directions that are only weakly constrained by the data, producing rapidly varying reconstructed histories. Several strategies have been developed to regularize flexible non-parametric reconstructions, including Gaussian Process (GP) methods and correlation priors on the reconstructed node values. A GP models the reconstructed function as a stochastic process whose correlations and smoothness are specified by a kernel, possibly with hyperparameters inferred from the data~\cite{Rasmussen2006Gaussian,Shafieloo:2012ht,Seikel:2012uu,Gerardi:2019obr,DESI:2025fii}. Alternatively, one can employ the correlation-prior approach, where the function is represented by a finite set of amplitudes, and a Gaussian prior with a prescribed covariance correlates their values at different times. This suppresses poorly constrained rapid variations while retaining a flexible description of the function~\cite{Crittenden:2011aa,Zhao:2012aw,Zhao:2017cud,Gerardi:2019obr}.

In this work, we adopt the latter approach and use a theory-informed correlation prior derived from an ensemble of viable Horndeski models in Ref.~\cite{Espejo:2018hxa} and implemented for the joint reconstruction of $\Omega_{\rm X}$, $\mu$, and $\Sigma$ in Refs.~\cite{Pogosian:2021mcs,Raveri:2021dbu}. The covariance was constructed by Monte Carlo sampling the functions entering the effective field theory (EFT) description of Horndeski gravity. For each realization, the background and linear perturbations were evolved and models affected by ghost or gradient instabilities, or inconsistent with broad cosmological and local gravity constraints, were discarded. The resulting ensemble contained of order $10^4$ viable models and was used to estimate the correlations between different times within each function, as well as the cross-correlations among $\Omega_{\rm X}$, $\mu$, and $\Sigma$.

In the present analysis, we use the joint covariance $\mathbf{C}_{\rm H}$ derived from this ensemble as an additional Gaussian prior
\begin{equation}
    \chi^2_{\rm H}
    =
    (\mathbf{f}-\mathbf{f}_{\rm fid})^{\rm T}
    \mathbf{C}_{\rm H}^{-1}
    (\mathbf{f}-\mathbf{f}_{\rm fid})\,.
    \label{eq:horndeski_prior}
\end{equation}
so that the total effective chi-squared sampled by the MCMC is
\begin{equation}
    \chi^2
    =
    \chi^2_{\rm data}
    +
    \chi^2_{\rm H},
    \label{eq:total_chi2}
\end{equation}

Equation~\eqref{eq:horndeski_prior} shows that a Gaussian correlation prior must be defined relative to a fiducial history. Fixing this history in advance could influence the reconstruction in directions that are weakly constrained by the data. We therefore adopt the floating-fiducial prescription of Ref.~\cite{Crittenden:2011aa}, in which the fiducial history is determined from the sampled reconstruction itself.

At each point in the MCMC, we compute a local running average separately within each of the $\Omega_{\rm X}$, $\mu$, and $\Sigma$ sectors. For an interior node, the fiducial value is the average of that node and its two nearest neighbours on either side. Near the boundaries, a three-node average is used. Thus, the prior compares each sampled function with a locally smoothed version of itself rather than with an externally selected history.

Defining $\mathbf{S}$ as the block-diagonal matrix that computes the running average for all nodes, the fiducial vector and its residual can be expressed compactly as
\begin{equation}
    \mathbf{f}_{\rm fid}=\mathbf{S}\mathbf{f},
    \qquad
    \mathbf{f}-\mathbf{f}_{\rm fid}
    =(\mathbf{I}-\mathbf{S})\mathbf{f}.
    \label{eq:floating_fiducial}
\end{equation}
Substituting this residual into Eq.~\eqref{eq:horndeski_prior} gives
\begin{equation}
    \chi^2_{\rm H}
    =
    \mathbf{f}^{\rm T}\mathbf{P}_{\rm eff}\mathbf{f},
    \qquad
    \mathbf{P}_{\rm eff}
    \equiv
    (\mathbf{I}-\mathbf{S})^{\rm T}
    \mathbf{C}_{\rm H}^{-1}
    (\mathbf{I}-\mathbf{S}),
    \label{eq:effective_prior_precision}
\end{equation}
where $\mathbf{P}_{\rm eff}$ is the effective Gaussian precision after applying the running average.

The resulting effective prior primarily penalizes departures from local smoothness. Rapid variations between neighbouring nodes differ strongly from their local average and are therefore suppressed, whereas broad variations receive a weaker penalty and remain more responsive to the data.
Moreover, because the averaging operator leaves a constant function unchanged, $\mathbf{S}\mathbf{u}=\mathbf{u}$ for a constant vector $\mathbf{u}$, one has $(\mathbf{I}-\mathbf{S})\mathbf{u}=0$. The Gaussian smoothing prior therefore does not constrain the overall constant level of any of the three reconstructed functions. Its effective precision has one constant null direction in each sector and hence rank 30 in the joint 33-dimensional node space.

\subsection{Numerical implementation}
We use \texttt{MGCAMB}~\cite{Wang:2023tjj} for the evolution of the background and linear perturbations and its interface with  the MCMC sampler implemented in \texttt{Cobaya}~\cite{Torrado:2020dgo,Lewis:2002ah,Lewis:2013hha} to sample the parameter space. In addition to the reconstruction nodes, the sampled cosmological sector contains $\Omega_b h^2$, $\Omega_c h^2$, $H_0$, $\tau$, $\ln(10^{10}A_s)$, $n_s$, and the phenomenological CMB lensing amplitude $A_L$, together with the nuisance and calibration parameters associated with the likelihoods. We assume spatial flatness and fix the neutrino sector to one massive species with $\sum m_\nu=0.06\,{\rm eV}$.

For the LSS analysis, we use the MG extension of the DES Year 3 and CMB lensing likelihood machinery developed in Ref.~\cite{Ye:2026yqk}. Gravitational lensing auto-correlations are evaluated from Weyl--Weyl spectra, galaxy--lensing cross-correlations from matter--Weyl spectra, and the magnification contributions are computed using the corresponding spectra involving the Weyl potential. The DES Year 3 shear-ratio likelihood is also included in the analysis.
The modified interfaces are publicly available in the \texttt{cosmosis2cobaya} and \texttt{cosmosis\_mg\_public} repositories.\footnote{\url{https://github.com/genye00/cosmosis2cobaya} and \url{https://github.com/genye00/cosmosis_mg_public}.}

We model intrinsic alignments using the tidal-alignment and tidal-torquing (TATT) prescription and additionally perform the reconstruction using the nonlinear alignment (NLA) model. Both prescriptions are widely used in weak-lensing analyses~\cite{DES:2026fyc, DES:2026mkc}: NLA describes intrinsic alignments through a linear response to the tidal field, using the nonlinear matter power spectrum, while TATT adds tidal-torquing and density-weighting contributions, providing a more flexible description~\cite{Blazek:2017wbz}. Since changes in the intrinsic alignment contribution can be degenerate with modifications to the reconstructed gravitational response, we compare the two prescriptions to assess the sensitivity of our reconstruction to intrinsic-alignment modelling. Switching from TATT to NLA produces similar reconstructed histories and leaves our main conclusions unchanged. Details of the comparison are presented in Appendix~\ref{app:ia}.

Convergence is assessed using the Gelman--Rubin statistic~\cite{Gelman:1992zz}, requiring $R-1<0.01$.

\begin{figure*}
    \centering
    \includegraphics[width=0.8\linewidth]{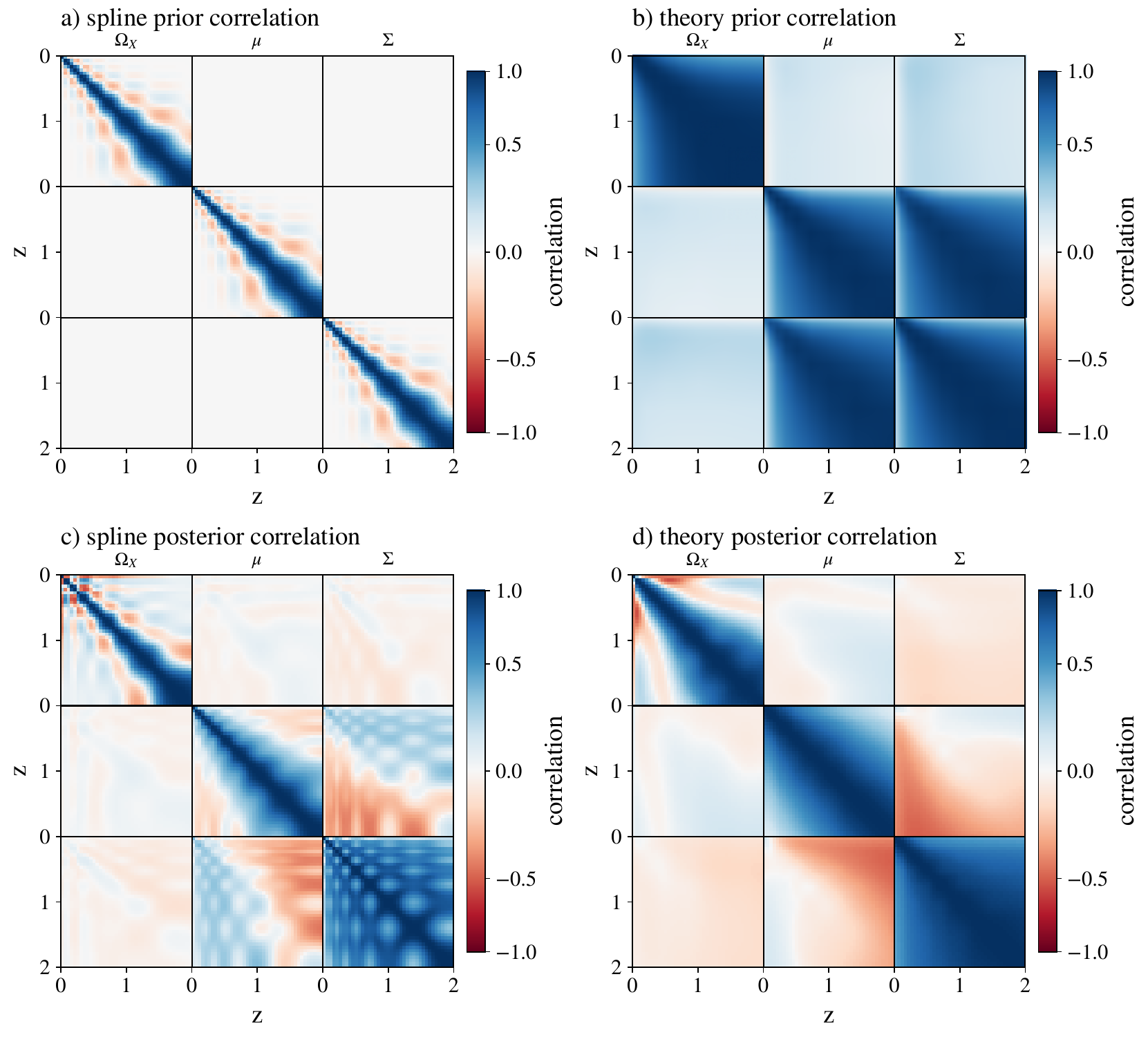}
    \caption{
    Correlation matrices for the joint reconstruction of $\Omega_{\rm X}$, $\mu$, and $\Sigma$. Panel (a) shows the implicit correlations between reconstructed function values induced by cubic-spline interpolation, which smoothly connects neighbouring reconstruction nodes. Panel (b) shows the correlation structure of the Horndeski prior, which correlates nodes both within and across the reconstructed functions and, in particular, produces a pronounced correlation between $\mu$ and $\Sigma$. Panel (c) shows the posterior correlations obtained from the Baseline data without the Horndeski prior, and therefore reflects correlations selected by the data together with those induced by the spline representation. Panel (d) shows the corresponding Baseline posterior correlations when the Horndeski prior is included.
    }
    \label{fig:correlations}
\end{figure*}
\subsection{Datasets}
We consider two principal dataset combinations. The first, denoted Baseline, combines primary CMB, CMB lensing, BAO, and supernova information. It contains the Planck low-$\ell$ temperature and polarization likelihoods~\cite{Planck:2019nip,Pagano:2019tci}, the ACT DR6 primary-CMB likelihood~\cite{AtacamaCosmologyTelescope:2025blo}, the Planck PR4 CMB lensing likelihood~\cite{Carron:2022eyg}, DESI DR2 BAO measurements~\cite{DESI:2025zpo,DESI:2025zgx}, and the uncalibrated Pantheon+ sample of Type Ia supernovae (SNe)~\cite{Scolnic:2021amr,Brout:2022vxf}.

The second combination, denoted Baseline+LSS, extends this analysis with late-time structure information. It includes the DES Year 3 (Y3) two-point measurements of galaxy clustering, cosmic shear, and galaxy--galaxy lensing ($3\times2$pt)~\cite{DES:2021wwk,DES:2021bpo}, their cross-correlations with CMB lensing from SPT and Planck~\cite{DES:2022qdz,DES:2022xxr,DES:2022urg}, and the SDSS DR16 BAO+RSD measurements for LRGs, ELGs, and QSOs~\cite{Alam:2020sor}. In this combination, the standalone Planck PR4 lensing likelihood used in Baseline is replaced by the combined ACT DR6+Planck+SPT-3G CMB lensing likelihood~\cite{ACT:2025qjh,ACT:2023kun,ACT:2023dou,ACT:2023ubw,SPT-3G:2024atg}.
To reduce sensitivity to uncertainties in nonlinear modelling, we adopt the DES Y3 linear scale cuts for the $3\times2$pt measurements~\cite{DES:2022ccp}. For DES--CMB-lensing cross-correlations, we adopt the published angular scale cuts corresponding to the linear galaxy-bias analysis~\cite{DES:2022qdz,DES:2022xxr}.
Both  dataset combinations are analysed with and without the Horndeski correlation prior.

\subsection{Statistical significance and constrained modes}
\label{statistical_analysis}
We characterize departures from $\Lambda$CDM both in the full reconstruction space and in combinations of nodes that are constrained by the data. The global significance is quantified using the signal-to-noise ratio (SNR), defined through the correlated distance between the posterior mean $\bar{\boldsymbol f}$ and the reference vector $\boldsymbol f_0$  corresponding to $\Lambda$CDM,
\begin{equation}
    \mathrm{SNR}_{\rm global}^{2}
    =
    (\bar{\boldsymbol f}-\boldsymbol f_0)^{\rm T}
    \mathbf{C}_{p}^{-1}
    (\bar{\boldsymbol f}-\boldsymbol f_0)\,,
    \label{eq:global_snr}
\end{equation}
where $\boldsymbol f$ denotes either the full reconstruction vector for the joint analysis or the corresponding subvector for an individual sector, and $\mathbf{C}_p$ is its posterior covariance. This statistic retains the correlations among the node amplitudes and is distinct from the significance assigned to any individual principal mode.

To determine which combinations of nodes are informed by the likelihood beyond the smoothing already supplied by the prior, we use covariant principal-component analysis (CPCA). We apply CPCA both to the full covariance matrix of the joint reconstruction and to the covariance submatrices associated with each reconstructed function. Here we show the CPCA eigenvalues for both decompositions, while restricting the analysis of mode displacements and redshift structure to the individual sectors. Combined-mode displacements and sector composition are studied in the companion paper~\cite{CompanionPRL}. Let $\mathbf{C}_{\Pi}$ and $\mathbf{C}_p$ denote, respectively, the reference-prior and posterior covariances defined on the same set of prior-constrained directions. We solve the generalized eigenvalue problem
\begin{equation}
    \mathbf{C}_{\Pi}\boldsymbol\Psi_a
    =
    R_a\,\mathbf{C}_p\boldsymbol\Psi_a\,,
    \label{eq:cpca_eigenproblem}
\end{equation}
where $\boldsymbol\Psi_a$ is the $a$th principal mode and
\begin{equation}
    R_a
    =
    \frac{\sigma_{\Pi,a}^{2}}
         {\sigma_{p,a}^{2}}
\end{equation}
measures the reduction in its variance relative to the prior. A narrow posterior mode is therefore not necessarily constrained by the data, since the prior may already have strongly reduced its variance. 
\begin{figure*}
    \centering
    \includegraphics[width=0.8\linewidth]{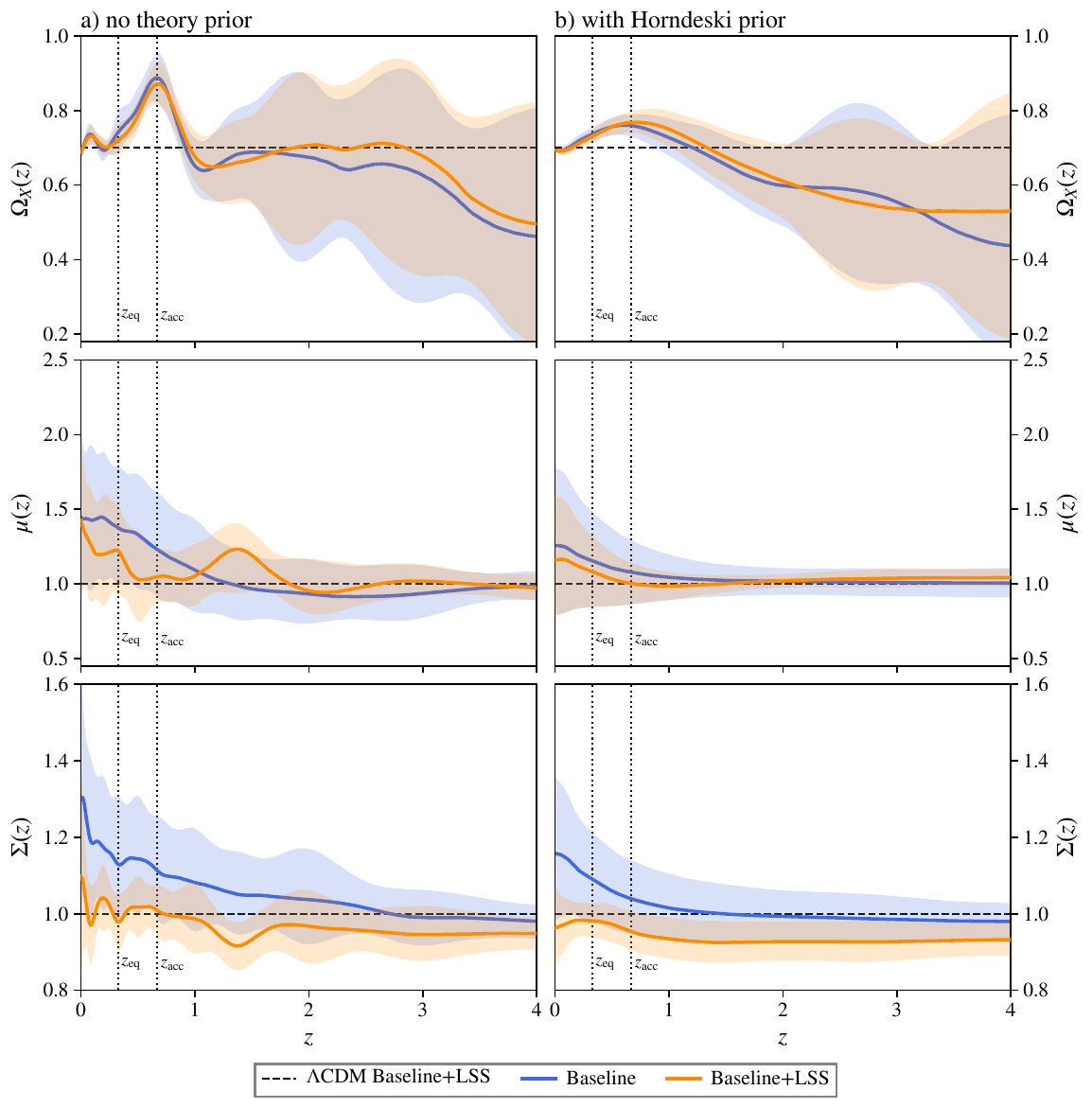}
    \caption{
    Joint reconstruction of $\Omega_{\rm X}$, $\mu$, and $\Sigma$ for the Baseline and Baseline+LSS data combinations, without and with the Horndeski prior. Solid curves indicate posterior medians and shaded regions the central 68\% credible intervals at each redshift.
    }
    \label{fig:reconstruction}
\end{figure*}
For the running-average Gaussian prior, the comparison is performed after projecting out the constant null direction in each reconstructed sector. This leaves 10 prior-constrained directions for each function and 30 directions in the joint reconstruction space. The SNR of an individual mode is defined as
\begin{equation}
    \mathrm{SNR}_a
    =
    \frac{
    \left|
    \boldsymbol\Psi_a^{\rm T}
    (\bar{\boldsymbol f}-\boldsymbol f_0)
    \right|
    }{
    \sqrt{
    \boldsymbol\Psi_a^{\rm T}
    \mathbf{C}_p
    \boldsymbol\Psi_a
    }
    }\,.
    \label{eq:modal_snr}
\end{equation}
This quantity measures the displacement of a single CPCA mode from its $\Lambda$CDM value.

\section{Results}
\label{sec:results}
We now present the results of the reconstruction. We first examine the reconstructed background and perturbation histories and assess their robustness to the inclusion of the Horndeski correlation prior and of LSS  information. We then use the CPCA decomposition to distinguish directions that are genuinely constrained by the data from those dominated by the prior, and quantify their displacement from the $\Lambda$CDM expectation. Finally, we examine the impact of the reconstruction on the standard cosmological parameters.

\subsection{Joint reconstruction of the background and gravitational sector}
\label{subsec:joint_rec}
In Fig.~\ref{fig:reconstruction}, we present the joint reconstruction of $\Omega_X$, $\mu$, and $\Sigma$ for the Baseline and Baseline+LSS data combinations, both with and without the Horndeski correlation prior. A first notable effect of the theory prior is the substantial suppression of the oscillatory features present in the data-only reconstructions. These rapid variations arise from the interplay between the flexibility of the cubic-spline parametrization, the adopted binning scheme, and weakly constrained directions in the reconstruction space. They should therefore not necessarily be interpreted as features selected by the data. This interpretation is supported by the correlation matrices shown in Fig.~\ref{fig:correlations}: the data-only correlations exhibit a characteristic ring-like substructure, with a frequency closely matching that associated with the cubic-spline interpolation. Once the Horndeski correlation prior is included, this high-frequency structure is strongly suppressed, resulting in considerably smoother posterior correlations and reconstructed histories.

Returning to Fig.~\ref{fig:reconstruction}, and focusing first on the Baseline reconstruction, the most prominent feature emerges in the background sector. The reconstructed $\Omega_X$ exhibits a broad low redshift maximum around $z\simeq 0.6$--$0.7$, which persists after the inclusion of the correlation prior, although with a reduced amplitude (see Table~\ref{tab:omega_x_peak_joint}). The perturbation sector presents a less clear-cut picture. Both $\mu$ and $\Sigma$, while remaining broadly compatible with their GR values, exhibit mild departures from unity over parts of the reconstructed redshift range. In particular, both functions show a tendency towards values above unity at low redshift, approximately coinciding with the range in which $\Omega_X$ rises above $\Omega_\Lambda$ and reaches its maximum.
Adding the LSS likelihoods leaves the main feature in
$\Omega_X$ qualitatively unchanged, suggesting that the
reconstructed low-redshift maximum is not primarily driven
by the additional clustering and lensing information.
The main impact of LSS is instead visible in the
perturbation sector, where the additional data substantially
reduce the allowed freedom in $\mu$ and $\Sigma$, bringing
their reconstructed evolutions closer to the GR expectation
over much of the redshift range. A mild tendency towards
$\mu>1$ persists at low redshift, whereas $\Sigma$ shifts
towards values below unity, particularly when the
Horndeski prior is included. We discuss the implications
of this behaviour further in Sec.~\ref{subsec:cosm_pars}.

The statistical relevance of these departures cannot,
however, be inferred directly from the marginalized
credible intervals in Fig.~\ref{fig:reconstruction}.
We return to this question in the following subsection,
where we quantify departures from the $\Lambda$CDM
expectation using the full posterior covariance and
the CPCA modes.

Let us conclude this subsection with a remark on the reconstructed background evolution in more detail. Its qualitative behaviour is consistent with previous indications of an evolving dark-energy sector, in particular those reported by the DESI Collaboration~\cite{DESI:2025fii,DESI:2024aqx}, although here it emerges within the considerably broader framework of a joint, non-parametric reconstruction of the background and perturbation sectors. In Sec.~\ref{sec:background_recon}, we dive deeper on the bump, showing that its prominence, location in redshift, and inferred statistical relevance depend to some extent on the particular combination of cosmological data entering the reconstruction.
\begin{table}
\caption{Location of the low-redshift maximum of $\Omega_{\rm X}$. The interval is the central 68\% credible interval (CI) of the peak locations. }
\label{tab:omega_x_peak_joint}
\begingroup
\setlength{\tabcolsep}{5pt}
\begin{tabular}{lcc}
\toprule
Analysis & $z_{\rm peak}$ & 68\% CI \\
\midrule
Baseline, no theory prior & 0.668 & $[0.574,\,0.701]$ \\
Baseline+LSS, no theory prior & 0.680 & $[0.616,\,0.711]$ \\
Baseline, Horndeski prior & 0.620 & $[0.476,\,0.898]$ \\
Baseline+LSS, Horndeski prior & 0.721 & $[0.555,\,0.941]$ \\
\bottomrule
\end{tabular}
\endgroup
\end{table}
\subsection{Departures from $\Lambda$CDM and data-constrained modes}
\label{subsec:constrained_modes}
We shall now quantify departures from the $\Lambda$CDM reference, and investigate which combinations of nodes are measured by the data over the reference prior.
Table~\ref{tab:significance_of_detection} addresses the first question through the correlated posterior distances from the $\Lambda$CDM reference for each reconstructed sector. The background sector shows the largest displacement from the $\Lambda$CDM reference, while the corresponding distances in $\mu$ and $\Sigma$ are smaller. The Horndeski prior reduces the background displacement by suppressing rapid temporal structure, but leaves the qualitative hierarchy unchanged.  We stress that ${\rm SNR}_{\rm global}$ is a multidimensional posterior distance and is not, by itself, a Gaussian-tail significance or a model-selection statistic. Accordingly, we use it to compare the relative size of the reconstructed departures rather than to quote an $N\sigma$ rejection of $\Lambda$CDM.

\begin{figure}
    \includegraphics[width=\columnwidth]{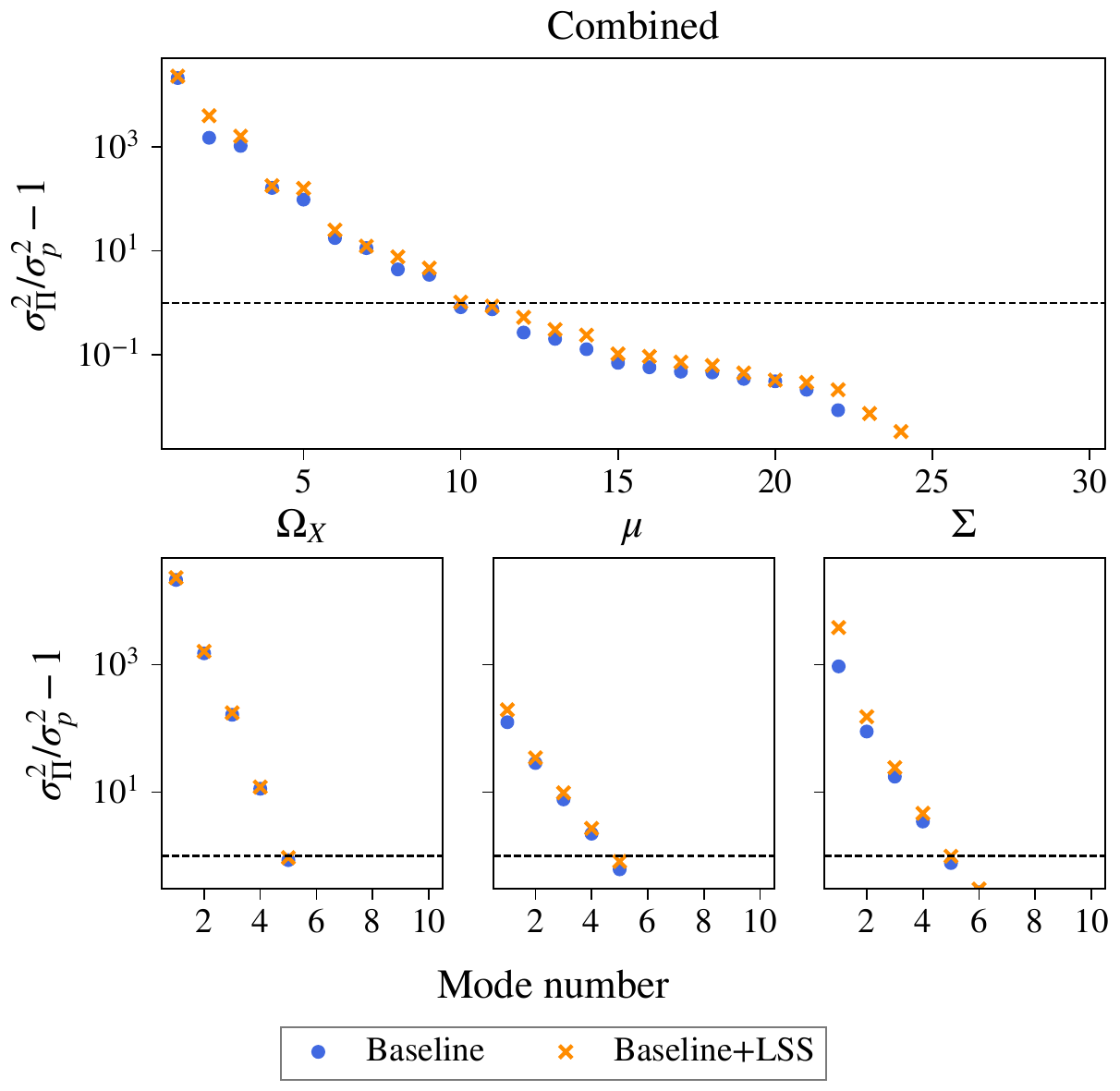}
    \caption{
    CPCA eigenvalues for the combined reconstruction and for the individual $\Omega_{\rm X}$, $\mu$, and $\Sigma$ sectors.
    }
    \label{fig:eigenvalues}
\end{figure}

We next investigate which combinations of reconstruction nodes are constrained by the data beyond the information supplied by the correlation prior. Figure~\ref{fig:eigenvalues} shows the corresponding CPCA variance ratios discussed in Sec.~\ref{statistical_analysis}. Several modes, roughly $4$, in each of the three sectors have posterior variances substantially smaller than those supplied by the reference prior, demonstrating that \emph{current data constrain multiple independent combinations of background and linear perturbations}. In particular, the leading $\Omega_{\rm X}$ mode has a large CPCA variance ratio, indicating that this combination of background nodes is strongly constrained by the data over the reference prior.

\begin{table}
\caption{
Global posterior distances ${\rm SNR}_{\rm global}$ from the $\Lambda$CDM reference.
}
\label{tab:significance_of_detection}
\begin{ruledtabular}
\begin{tabular}{cccc}
SNR & $\Omega_{\rm X}$ & $\mu$ & $\Sigma$ \\
\hline
\multicolumn{4}{l}{no theory prior} \\
\hline
Baseline & 4.0 & 1.6 & 1.7 \\
Baseline+LSS & 3.8 & 1.5 & 2.2 \\
\hline
\multicolumn{4}{l}{with Horndeski prior} \\
\hline
Baseline & 2.7 & 0.7 & 1.5 \\
Baseline+LSS & 2.8 & 1.1 & 1.7 \\
\end{tabular}
\end{ruledtabular}
\end{table}

\begin{figure}
    \includegraphics[width=0.9\columnwidth]{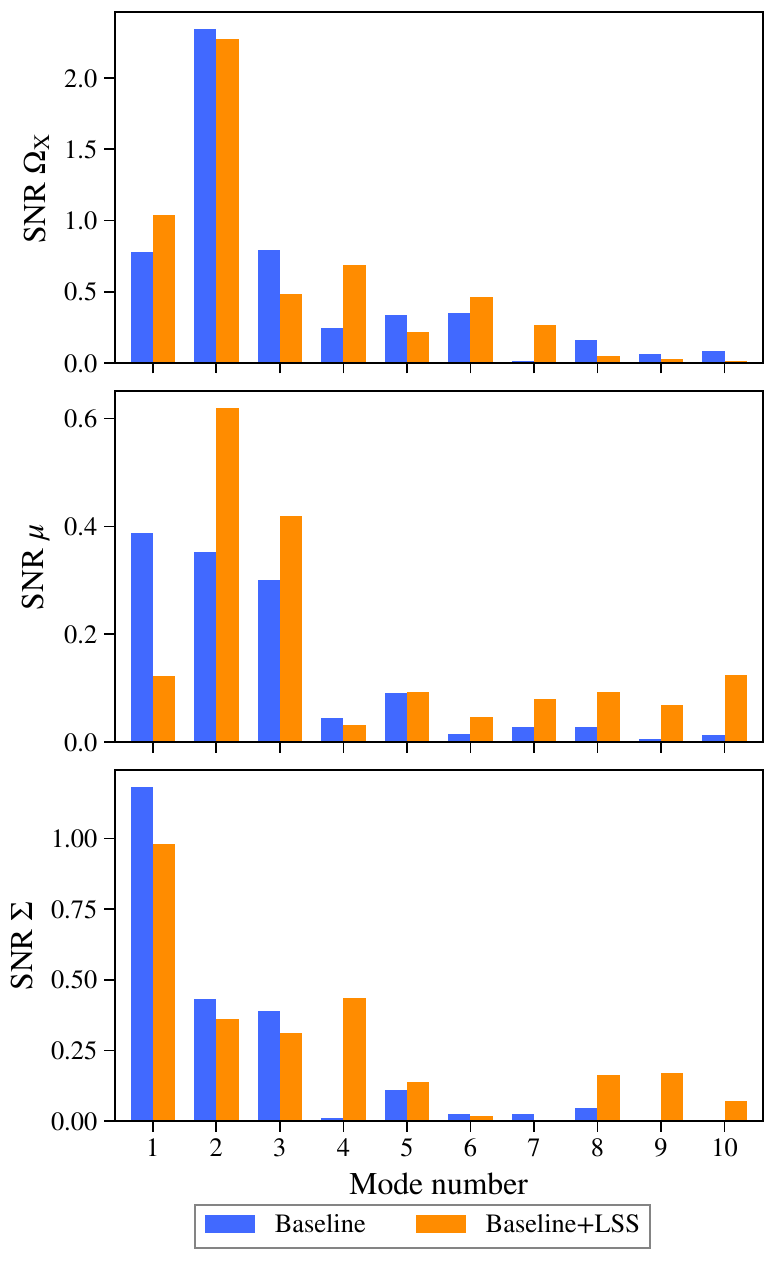}
    \caption{
    Signal-to-noise ratios of the individual-sector CPCA modes from the joint reconstruction. From top to bottom, the panels show the $\Omega_{\rm X}$, $\mu$, and $\Sigma$ sectors.
    }
    \label{fig:SNR}
\end{figure}

Having identified the combinations constrained by the data, we now examine their displacement from the $\Lambda$CDM reference through the individual-mode signal-to-noise ratios shown in the three panels of Fig.~\ref{fig:SNR}. The clearest departure is found in the background sector, where the second $\Omega_{\rm X}$ mode has ${\rm SNR}\simeq2.3$ for both Baseline and Baseline+LSS. The remaining background modes have substantially smaller displacements. By contrast, the individually constrained $\mu$ modes remain below unit SNR, while only the leading $\Sigma$ mode approaches order-unity displacement. In other words, although the data constrain several combinations in all three sectors, the largest individual-mode departure from $\Lambda$CDM occurs in $\Omega_{\rm X}$.

\begin{figure}
    \includegraphics[width=0.9\columnwidth]{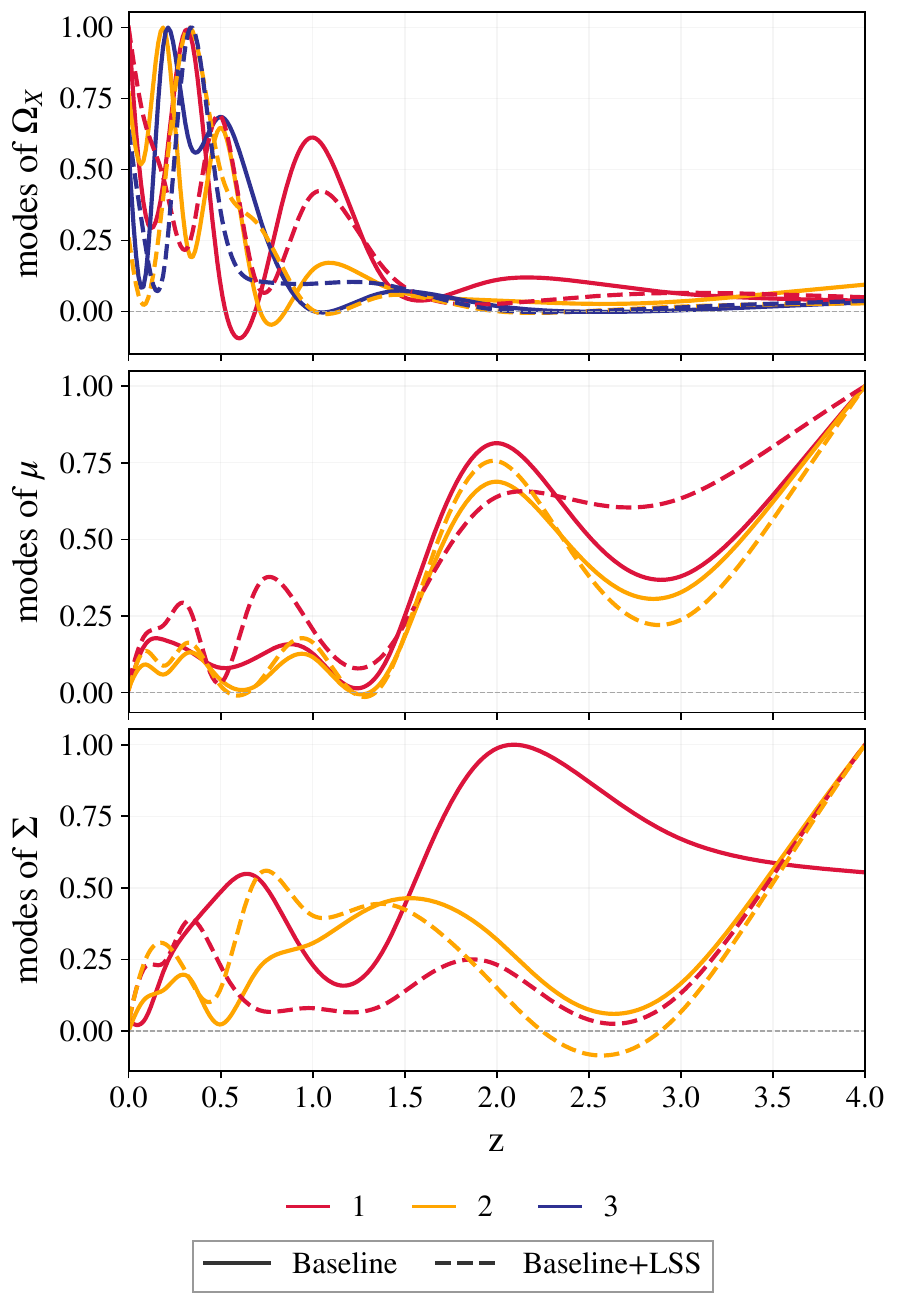}
    \caption{
    Shapes of the amplitudes of the individual CPCA modes of $\Omega_{\rm X}$, $\mu$ and $\Sigma$ for Baseline (solid lines) and Baseline+LSS (dashed lines). Curves are obtained by interpolating the absolute values of the node weights and are rescaled so that the maximum amplitude of each curve is unity.
    }
    \label{fig:eigenmodes}
\end{figure}

The eigenvectors also provide information on the
redshift structure of the combinations constrained
by the data. Figure~\ref{fig:eigenmodes} shows the
absolute values of the individual-sector CPCA
eigenvector components, interpolated between
reconstruction nodes and normalized to unit
maximum amplitude. These curves indicate the
redshift regions carrying the largest weights
in each mode, rather than observational window
functions or direct measures of sensitivity.
The leading $\Omega_{\rm X}$ modes are
predominantly weighted towards low redshift,
whereas the $\mu$ and $\Sigma$ modes generally
extend over a broader interval. Adding LSS
information mainly modifies the gravitational
modes, while the background modes retain
their strongest weights at low redshift.

Taken together, the CPCA and displacement analyses indicate
that present data have greater sensitivity to the background expansion, and an appreciable constraining power on the
perturbation sector, particularly after the inclusion of LSS
information. The combined-mode analysis presented in our companion Letter provides a complementary view of this hierarchy, including the relative contributions of background and perturbation variables to the constrained directions. While
this hierarchy may change with the next generation of LSS
surveys, departures from the $\Lambda$CDM expectation
are currently more significant in the background sector
than in linear gravitational dynamics.

This result complements theory-driven studies in which
specific scalar-tensor realizations impose relations between
phantom-crossing background evolution and modified growth~\cite{Garcia-Garcia:2026nzy}.
Here, instead, the background and perturbation sectors are
allowed to vary independently, apart from the mild correlations
introduced by the theory prior. Importantly, the reconstructed
$\mu$ and $\Sigma$ exhibit suggestive trends in the
gravitational response, which future Stage-IV LSS observations
will be able to test with greater precision, potentially
revealing whether they share a common physical origin with
the reconstructed background evolution. We explore these
features further in Sec.~\ref{subsec:interpretation}, where
we examine the gravitational slip and interpret the
reconstructed functions within the framework of
scalar-tensor theories.

\subsection{Impact on standard cosmological parameters}
\label{subsec:cosm_pars}
\begin{figure}[htbp]
    \centering
    \includegraphics[width=\columnwidth]{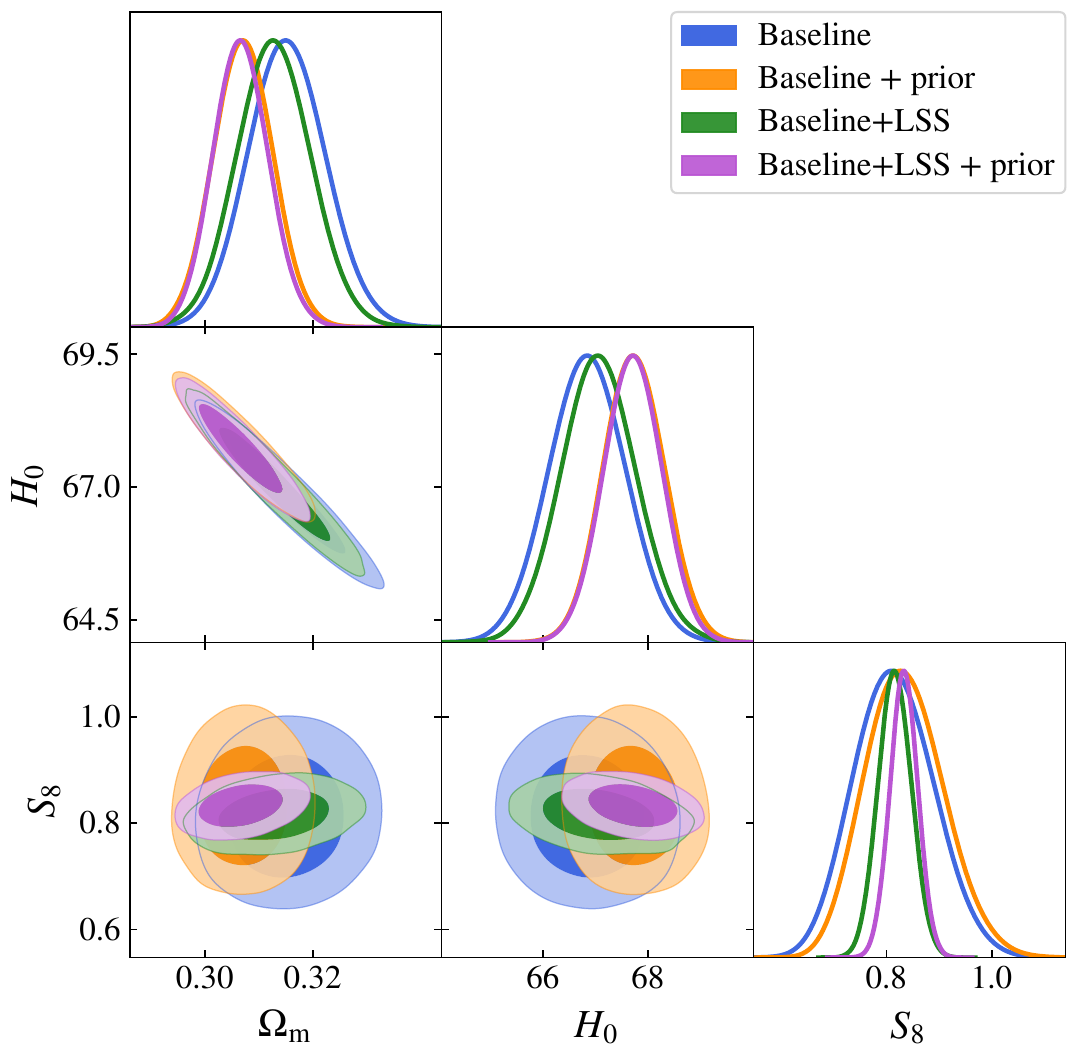}
    \medskip
    \includegraphics[width=\columnwidth]{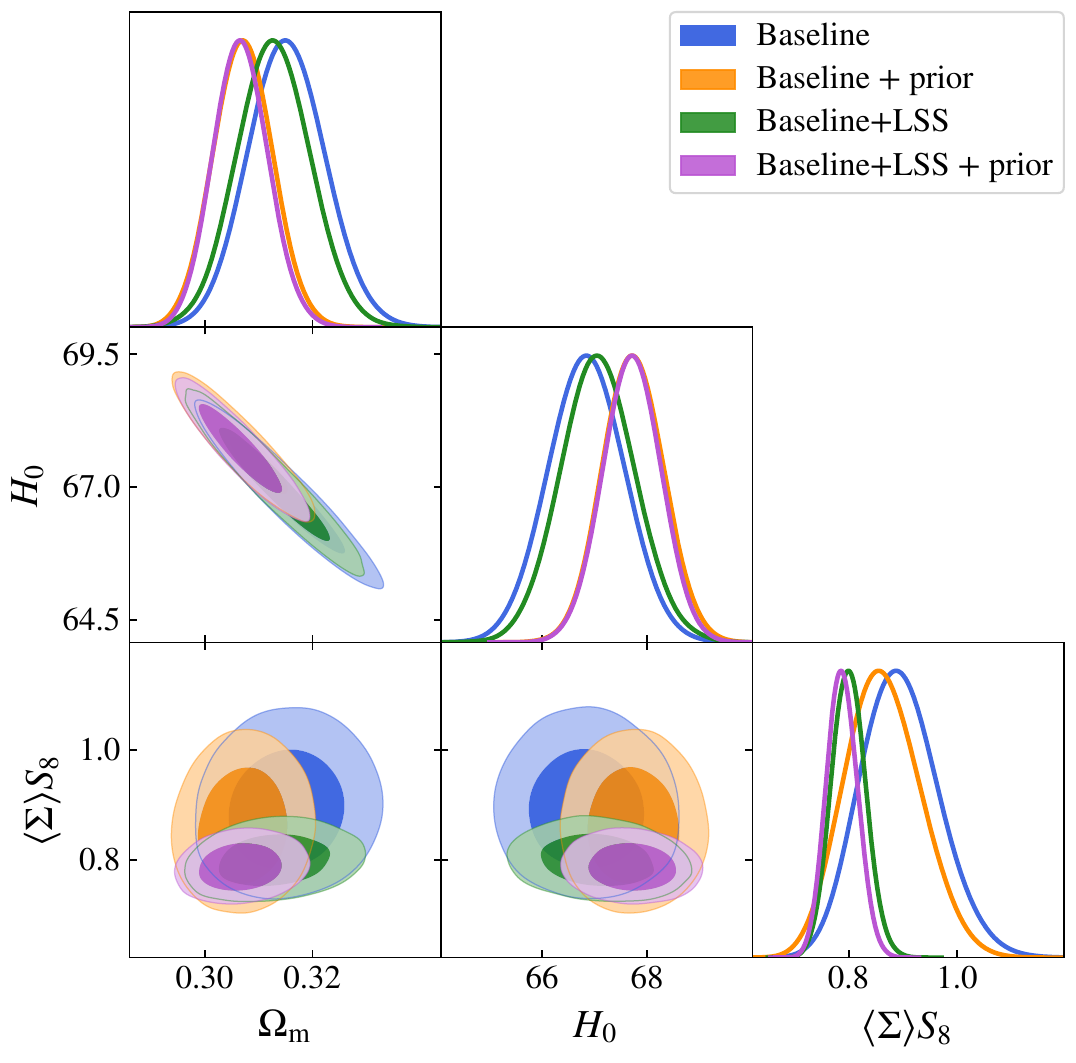}
    \caption{
    \emph{Upper panel}: The 68\% and 95\% credible contours
    for $\Omega_{\rm m}$, $H_0$, and $S_8$, from the Baseline
    and Baseline+LSS data with and without the Horndeski
    prior in the joint reconstruction.
    \emph{Lower panel}: The same, with
    $\langle\Sigma\rangle S_8$ replacing $S_8$.
    }
    \label{fig:tr_plot_joint}
\end{figure}

The upper panel of Fig.~\ref{fig:tr_plot_joint} shows the marginalized constraints on $\Omega_m$, $H_0$, and $S_8$ for the Baseline and Baseline+LSS combinations, with and without the Horndeski correlation prior. The inclusion of the prior leads to a moderate tightening of the constraints on the background parameters, accompanied by a shift towards lower $\Omega_m$ and higher $H_0$, largely along the existing degeneracy direction. The inclusion of LSS information produces a similar, although generally smaller, shift. Interestingly, the inferred value of $S_8$ remains relatively stable across the different analyses.

The lower panel of Fig.~\ref{fig:tr_plot_joint} shows the corresponding constraints when $S_8$ is replaced by $\langle\Sigma\rangle S_8$, where $\langle\Sigma\rangle$ denotes the average of the reconstructed lensing modification over $0<z<2$, approximately corresponding to the redshift range probed by DES Y3 galaxy weak lensing. This combination provides a simple proxy for the degeneracy between the amplitude of matter fluctuations and modifications to the Weyl potential entering weak-lensing observables. While the effective sensitivity of galaxy lensing is concentrated at relatively low redshifts, CMB lensing probes a broader range, peaking around $z\simeq 2$. The two therefore constrain different redshift-weighted combinations of $\Sigma(z)$, with their cross-correlations providing complementary information.

Unlike $S_8$, the inferred $\langle\Sigma\rangle S_8$ exhibits a noticeable downward shift and a substantial tightening of its constraints upon inclusion of LSS information. This behaviour is consistent with the reconstructed evolution of $\Sigma$ in Fig.~\ref{fig:reconstruction}, which shifts towards values below unity in the Baseline+LSS analysis. It suggests that the additional LSS information favours a reduced effective lensing amplitude, which can be accommodated through a lower $\Sigma$ without requiring a corresponding shift in the inferred amplitude of matter fluctuations $S_8$. The different redshift sensitivities of galaxy and CMB lensing, together with the complementary constraints from galaxy clustering, may contribute to this behaviour.

\begin{figure*}
    \centering
    \includegraphics[width=0.9\linewidth]{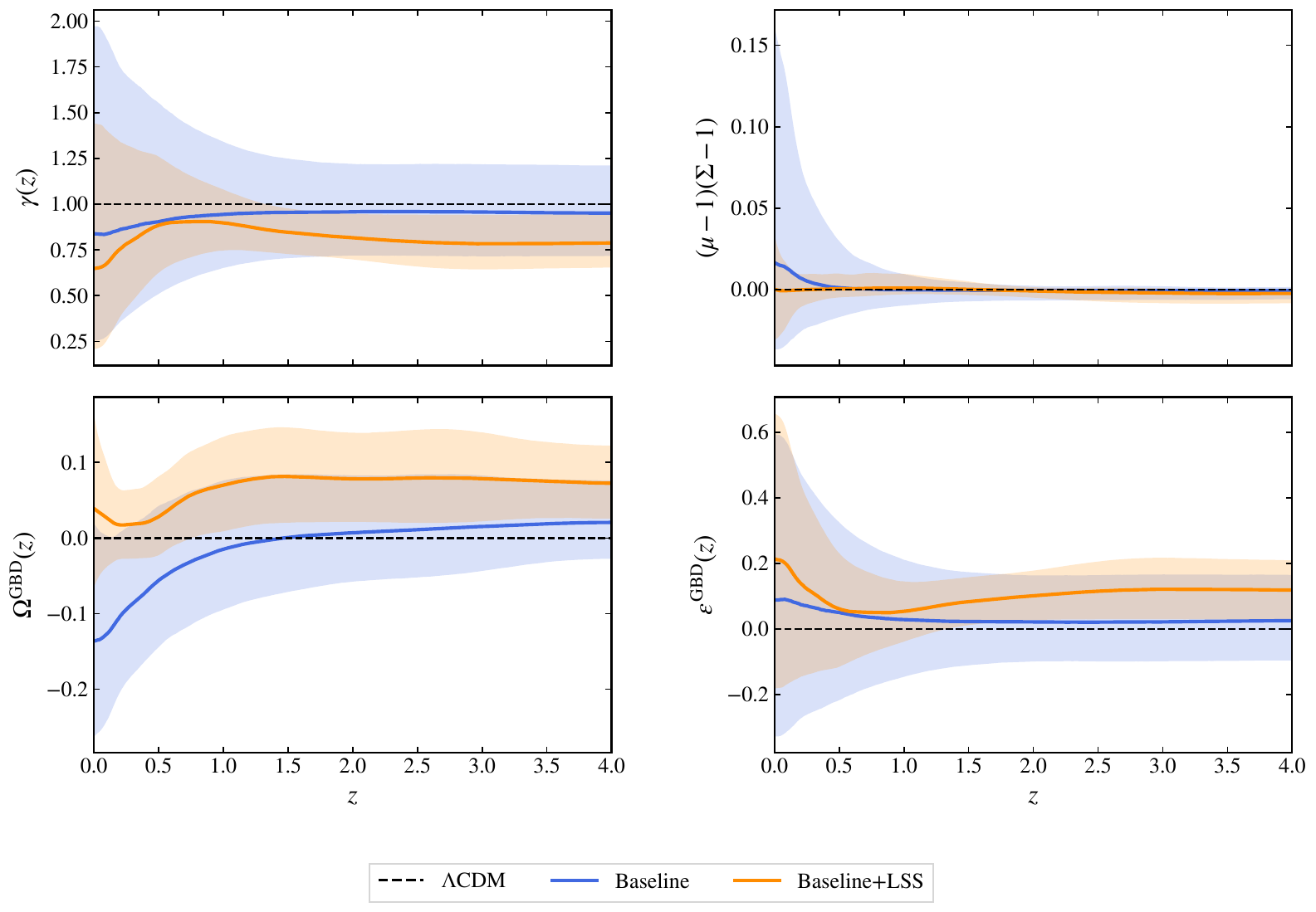}
    \caption{
    Derived gravity functions from the joint posterior of the reconstructed functions $\mu$ and $\Sigma$, for the Baseline and Baseline+LSS data combinations, both including the Horndeski correlation prior. The panels show the gravitational slip $\gamma=2\Sigma/\mu-1$, the combination $(\mu-1)(\Sigma-1)$, the conformal factor $\Omega^{\rm GBD}=\Sigma^{-1}-1$ and the fifth force contribution $\varepsilon^{\rm GBD}=\mu/\Sigma-1$. 
    }
    \label{fig:derived_functions}
\end{figure*}
\subsection{Interpretation within Horndeski}
\label{subsec:interpretation}
It is interesting to investigate whether the reconstructed gravitational functions exhibit features characteristic of particular classes of scalar-tensor theories. To this end, we consider several combinations of the reconstructed $\mu$ and $\Sigma$, shown in Fig.~\ref{fig:derived_functions}, which allow us to assess their consistency with theoretical expectations and explore a possible interpretation within Generalized Brans--Dicke (GBD) models.

We begin with the gravitational slip, defined as $\gamma\equiv\Phi/\Psi=2\Sigma/\mu-1$. In scalar-tensor theories featuring an attractive scalar-mediated fifth force, the quasi-static limit typically predicts $\gamma<1$. As shown in the upper-left panel of Fig.~\ref{fig:derived_functions}, both reconstructions exhibit a mild preference for $\gamma<1$, particularly when LSS information is included, although the reconstructed evolution remains broadly compatible with the GR expectation, $\gamma=1$.

Next, we consider the combination $(\mu-1)(\Sigma-1)$, which was used in~\cite{Pogosian:2016pwr} to formulate a conjectured consistency relation for Horndeski theories, $(\mu-1)(\Sigma-1)\geq 0$, under the assumptions discussed therein. The upper-right panel of Fig.~\ref{fig:derived_functions} shows that the reconstructed combination remains close to zero and is broadly compatible with the conjectured relation. We have verified that the qualitative behavior of both this combination and the gravitational slip is similar in reconstructions performed without the Horndeski correlation prior. This is relevant because, although the prior primarily suppresses rapidly varying histories, it also introduces theoretically motivated correlations between the reconstructed gravitational functions. Our results therefore reveal no compelling evidence for a violation of the conjectured relation, while not constituting an independent confirmation of it.

The tendency towards a reduced effective lensing response identified in the previous subsection can also be interpreted, within GBD models, as a positive non-minimal coupling that partially compensates for the scalar-mediated fifth force. We therefore explore an interpretation of the reconstructed functions within GBD theories. In the quasi-static regime, their gravitational response can be expressed as~\cite{Pogosian:2016pwr}
\begin{equation}
\label{eq:GBD_QS}
\mu=\frac{1}{1+\Omega^{\rm GBD}}\left(1+\varepsilon^{\rm GBD}\right)\,,\qquad
\Sigma=\frac{1}{1+\Omega^{\rm GBD}}\,,
\end{equation}
where $\Omega^{\rm GBD}$ parametrizes the non-minimal coupling, or equivalently the modification of the effective Planck mass, while $\varepsilon^{\rm GBD}$ describes the additional scalar-mediated fifth-force contribution. These relations allow us to derive
\begin{equation}
\label{eq:GBD_derived}
\Omega^{\rm GBD}=\Sigma^{-1}-1\,,\qquad
\varepsilon^{\rm GBD}=\frac{\mu}{\Sigma}-1\,.
\end{equation}
The corresponding reconstructions are shown in the lower panels of Fig.~\ref{fig:derived_functions}. Interestingly, $\varepsilon^{\rm GBD}$ exhibits a preference for positive values in both dataset combinations, consistent with an attractive scalar fifth force. The inferred non-minimal coupling, on the other hand, is more sensitive to the inclusion of LSS information. While Baseline tends to favor negative $\Omega^{\rm GBD}$, the addition of LSS shifts the reconstruction towards positive values, reflecting the corresponding downward shift in $\Sigma$.

This behavior highlights the distinct contributions of the non-minimal coupling and the fifth force to the gravitational response. In particular, a positive $\Omega^{\rm GBD}$ suppresses the effective gravitational coupling, potentially compensating for the enhancement associated with a positive $\varepsilon^{\rm GBD}$. Indeed, from Eq.~\eqref{eq:GBD_QS} one obtains
\begin{equation}
\mu-1=\frac{\varepsilon^{\rm GBD}-\Omega^{\rm GBD}}{1+\Omega^{\rm GBD}}\,,
\end{equation}
showing that $\mu$ can remain close to its GR value even when the individual contributions are nonzero. The reconstructed functions therefore admit an interesting phenomenological interpretation in terms of competing modifications to the gravitational interaction. We stress, however, that this mapping is indicative: establishing whether the reconstructed background and perturbation histories can be simultaneously realized within a viable GBD model would require a dedicated theoretical reconstruction and stability analysis.

\section{Auxiliary background reconstructions}
\label{sec:background_recon}
To further investigate the origin and robustness of the reconstructed expansion history, we additionally perform four prior-assisted reconstructions in which only $\Omega_{\rm X}(a)$ is varied while $\mu=\Sigma=1$. These analyses allow
us to examine the role of the different cosmological
datasets in shaping the reconstructed background evolution
and to assess whether its main features persist when
the additional freedom in the perturbation sector is removed.

As in the joint analysis, the first ten $\Omega_{\rm X}$ nodes are sampled and the present-day node is again fixed by flatness.  All four reconstructions
include the Horndeski correlation prior, restricted to
the background sector. More precisely, we adopt the
conditional $\Omega_X$ block obtained by fixing the
gravitational sector in the prior covariance matrix,
rather than the covariance marginalized over $\mu$
and $\Sigma$. With the same local running-average
prescription employed in the joint reconstruction,
the resulting effective Gaussian precision has rank $10$. 
\begin{figure}
    \centering
    \includegraphics[width=\columnwidth]{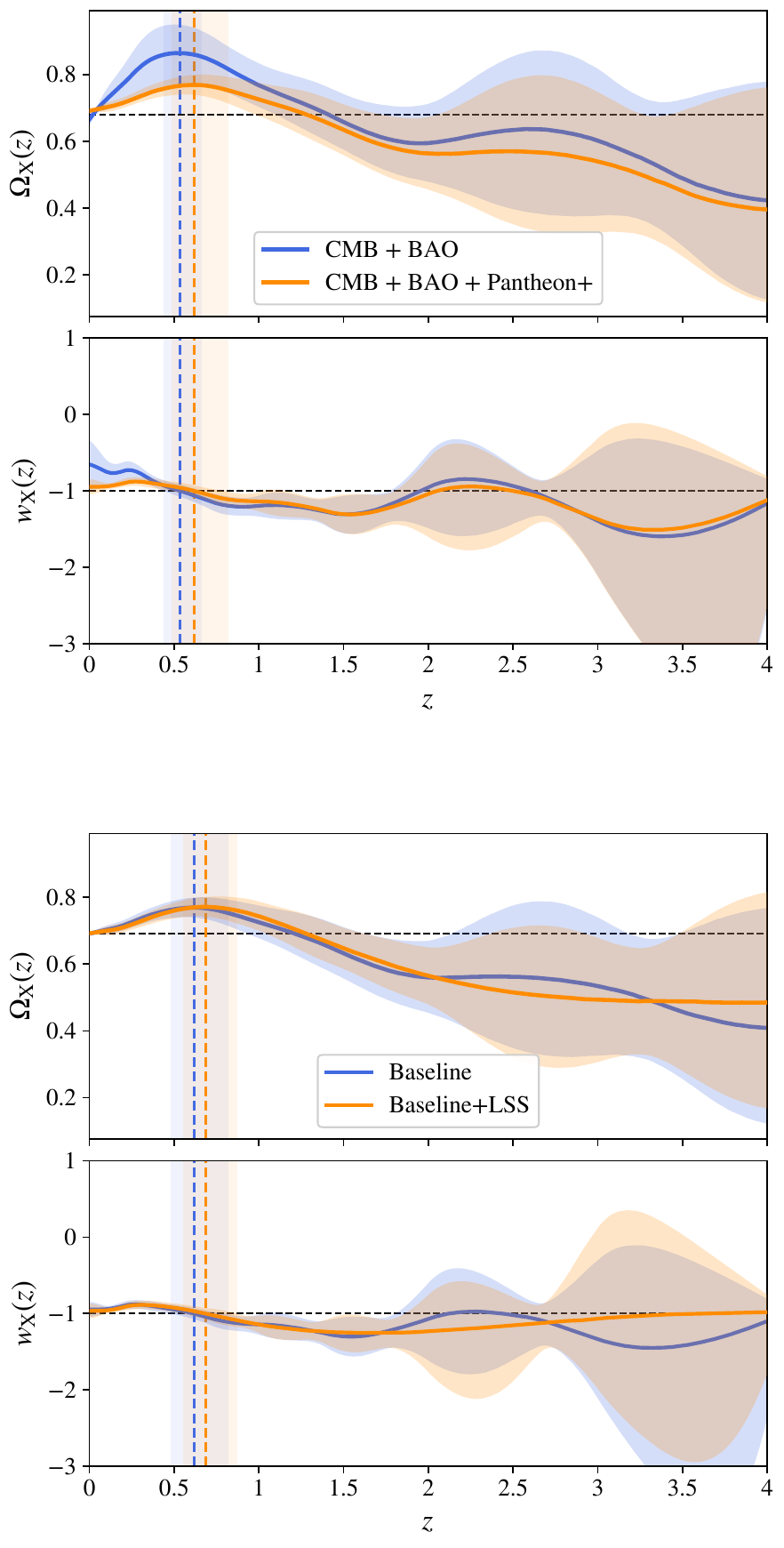}
    \caption{
    Reconstruction of $\Omega_{\rm X}$ and the corresponding derived effective equation of state $w_{\rm X}$ for the CMB+BAO, CMB+BAO+Pantheon+, Baseline and Baseline+LSS data combinations. In these analyses, $\mu=\Sigma=1$, and the Horndeski correlation prior is included. The horizontal dashed lines indicate the corresponding $\Lambda$CDM values. The vertical dashed lines mark the median posterior redshift at which $\Omega_{\rm X}$ reaches its peak, equivalently corresponding to the associated $w_{\rm X}=-1$ crossing. The vertical shaded bands show the central $68\%$ credible intervals of this feature location.
    }
    \label{fig:reconstruction_omega_x_only_combined_ver}
\end{figure}

An important methodological distinction concerns the
treatment of perturbations. In the joint reconstruction, we adopt the standard phenomenological interpretation at the core of \texttt{MGCAMB}: $\Omega_{\rm X}(a)$ is an effective contribution to the background expansion, which need not correspond to a separately conserved DE fluid, while $\mu(a)$ and $\Sigma(a)$ capture the perturbative effects of the underlying new physics. No separate DE perturbations are therefore evolved, which preserves the model-agnostic nature of the reconstruction. In the $\Omega_{\rm X}$-only reconstructions, instead, the perturbation sector is fixed to GR, $\mu=\Sigma=1$, and the evolving $\Omega_{\rm X}$ is interpreted as an effective dynamical DE component whose perturbations must be consistently included. We evolve them with the parametrized post-Friedmann (PPF) prescription, which remains regular when the effective equation of state crosses the phantom divide. In principle on the sub-horizon scales that dominate the constraining power of our data, these perturbations have a negligible impact, so the two treatments can be compared directly at the level of the reconstructed background. Nevertheless, we opt for running also a Baseline and Baseline+LSS background-only reconstruction, for a more robust comparison. 

Motivated by the proposal that the preference for phantom crossing may originate primarily from the background constraints selected by CMB and BAO data~\cite{Ye:2025ark}, rather than being generated by supernovae~\cite{Shlivko:2026jxa}, we first compare CMB+DESI DR2 BAO with the same combination supplemented by Pantheon+. Neither analysis includes CMB lensing or LSS  likelihoods. Figure~\ref{fig:reconstruction_omega_x_only_combined_ver} shows that a non-monotonic $\Omega_{\rm X}$ is already selected by CMB+BAO: the reconstruction rises above its present-day value and reaches a maximum at $z_{\rm peak}=0.534^{+0.130}_{-0.095}$ before decreasing at higher redshift. Adding Pantheon+  changes the amplitude of the peak and shifts its median location to $z_{\rm peak}=0.618$, with a broader 68\% interval, as reported in Table~\ref{tab:omega_x_peak}. The appearance of
the maximum in the CMB+BAO reconstruction indicates
that this feature does not require the
inclusion of supernova data.

\begin{table}
\caption{Location of the low-redshift maximum of $\Omega_{\rm X}$ for the auxiliary background reconstructions. The interval is the central 68\% credible interval of the peak locations.}
\label{tab:omega_x_peak}
\begingroup
\setlength{\tabcolsep}{10pt}
\begin{tabular}{lcc}
\toprule
Analysis & $z_{\rm peak}$ & 68\% CI \\
\midrule
CMB+BAO & 0.534 & $[0.439,\,0.664]$ \\
CMB+BAO+SNe & 0.618 & $[0.482,\,0.819]$ \\
Baseline & 0.618 & $[0.482,\,0.819]$ \\
Baseline+LSS & 0.686 & $[0.551,\,0.875]$ \\
\bottomrule
\end{tabular}
\endgroup
\end{table}
The lower panels of Fig.~\ref{fig:reconstruction_omega_x_only_combined_ver} display the corresponding derived $w_{\rm X}(z)$.  The
reconstructed maximum therefore translates into a
crossing of the effective phantom divide, from
$w_X<-1$ at higher redshift to $w_X>-1$ at lower
redshift. We next repeat the $\Omega_X$-only reconstruction
using the Baseline and Baseline+LSS data combinations.
The characteristic low-redshift maximum persists,
with median locations remaining in the same broad
redshift interval. Comparing these results with the
joint reconstructions of Sec.~\ref{subsec:joint_rec}, we find that the
location of the maximum is comparatively stable
under changes in the dataset and reconstruction
framework, whereas its amplitude and detailed shape
show greater sensitivity to these choices. The persistence of its location across the four background-only and four joint reconstructions indicates that the existence and approximate redshift of the turnover are more robust than its detailed amplitude. 

\begin{figure}
    \includegraphics[width=\columnwidth]{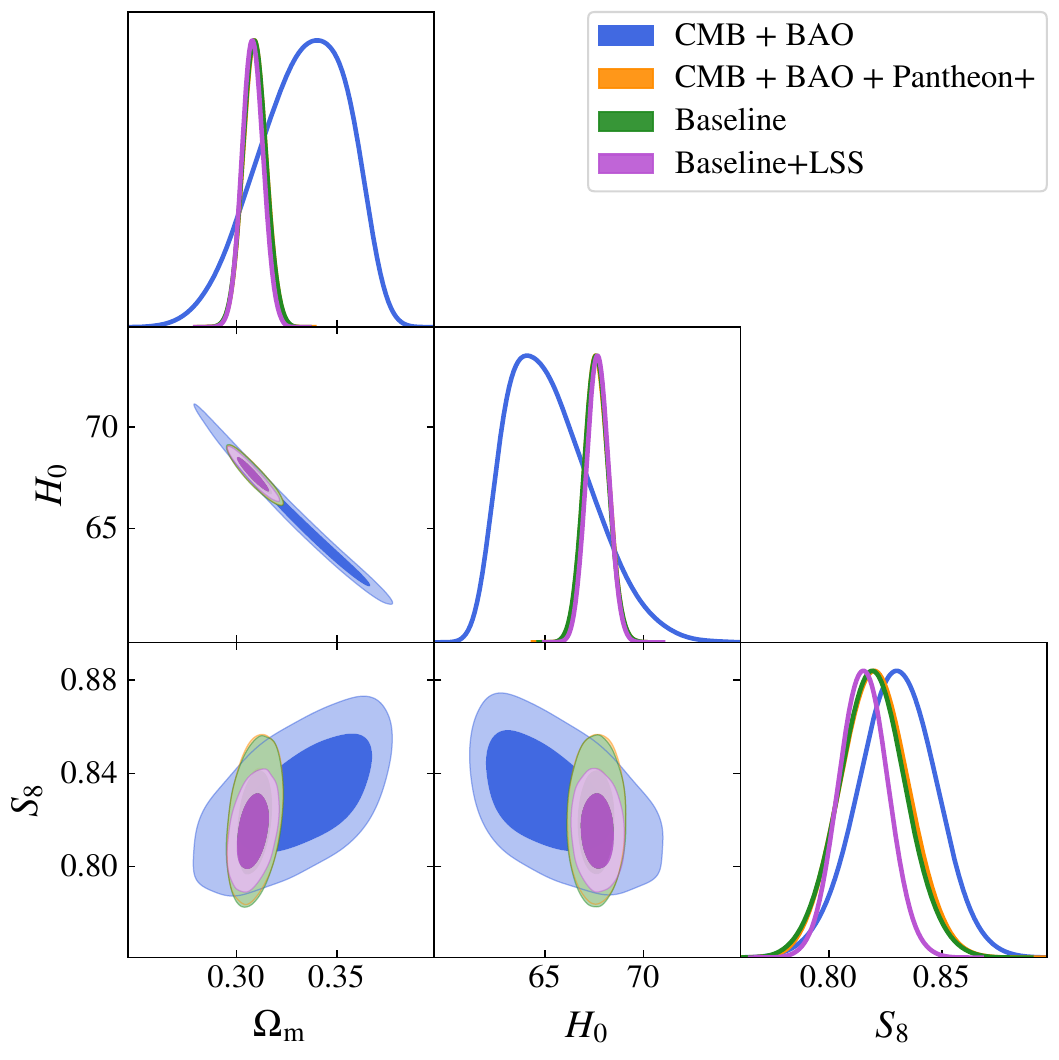}
    \caption{Marginalized constraints on $\Omega_{\rm m}$, $H_0$, and $S_8$ from the $\Omega_{\rm X}$-only  reconstructions. All four analyses fix $\mu=\Sigma=1$ and include the Horndeski correlation prior. 
    }
    \label{fig:triangle_omega_only}
\end{figure}

The corresponding marginalized constraints on the
standard cosmological parameters $\Omega_m$, $H_0$,
and $S_8$ are presented in Fig.~\ref{fig:triangle_omega_only}. Adding Pantheon+
to CMB+BAO produces a pronounced tightening of the
constraints, particularly in the $\Omega_m$--$H_0$
plane, accompanied by shifts in the inferred background
parameters. A further change is observed when moving
from CMB+BAO+Pantheon+ to the Baseline combination,
which additionally includes CMB lensing. In particular,
the degeneracy direction in the $\Omega_m$--$S_8$
plane rotates, reflecting the sensitivity of CMB
lensing to the amplitude of matter fluctuations.
The orientation of this degeneracy remains approximately
unchanged when the additional LSS likelihoods are
included in Baseline+LSS, while the corresponding
constraints undergo comparatively smaller changes.

Taken together, these auxiliary analyses indicate
that the non-monotonic evolution of $\Omega_X$ is
already present in the background constraints from
CMB and DESI DR2 BAO. Supernovae, CMB lensing,
and late-time LSS measurements refine the allowed
expansion history and affect the inferred cosmological
parameters, but do not qualitatively alter the
reconstructed low-redshift maximum. The agreement
between the joint and background-only reconstructions,
despite their different treatments of perturbations
and prior correlations, confirms the robustness of this feature.

\section{Conclusions}
\label{sec:conclusions}
Does dark energy evolve and cluster? Following the latest observational hints of a dynamical dark energy, we have asked ourselves whether dark energy in fact is dynamical and also clusters. In order to address this question,  we have performed a joint non-parametric reconstruction of late-time departures from $\Lambda$CDM in the background, clustering, and lensing sectors, described respectively by $\Omega_{\rm X}(a)$, $\mu(a)$, and $\Sigma(a)$. The analysis combines recent CMB and DESI DR2 BAO measurements with Pantheon+ Supernovae and, in the extended dataset, RSD from SDSS, DES Year 3 and CMB-lensing LSS information. By comparing reconstructions with and without a theory-motivated Horndeski correlation prior, and by using CPCA to separate data-constrained directions from prior-dominated ones, we have tested which features of the reconstructed histories are robust to the regularization of the high-dimensional node space.

The most persistent departure from the $\Lambda$CDM reference is found in the background sector. Across both the joint and $\Omega_{\rm X}$-only reconstructions, the effective DE density develops a maximum at $z\simeq0.5$--$0.7$. In terms of the derived effective equation of state, this corresponds to a crossing of $w_{\rm X}=-1$. The location of the feature is comparatively stable under changes in the dataset and under inclusion of the Horndeski prior, while its amplitude and detailed shape are more sensitive to these choices. This behaviour is qualitatively consistent with recent DESI non-parametric reconstructions of the expansion history, while here it is obtained in a framework that simultaneously allows departures in the perturbation sector.

The controlled background-only analyses further show that the basic non-monotonic evolution is already present in the CMB+DESI BAO combination. Adding Pantheon+ changes the inferred background parameters and the detailed shape and location of the turnover, but does not create the feature. CMB lensing and late-time structure information likewise leave its qualitative form intact. These results support an origin of the phantom-crossing behaviour primarily in the background constraints selected by CMB and BAO, with other low-redshift probes refining the allowed history.

The perturbation sector presents a different picture. Current data constrain several independent combinations of $\mu$ and $\Sigma$, particularly once LSS information is included, but the measured modes show smaller posterior displacements from the GR expectation than the leading background modes. In the Baseline reconstruction, we find mild tendencies towards $\mu>1$ and $\Sigma>1$ at low redshift, approximately coinciding with the range in which $\Omega_X$ rises above $\Omega_\Lambda$ and reaches its maximum.
With the inclusion of LSS, the tendency towards $\mu>1$ persists, although it becomes milder, whereas the reconstructed $\Sigma$ shifts below unity, indicating a reduced effective lensing response over the redshift range probed by the data. This tendency is somewhat more pronounced at higher redshifts, where CMB lensing provides substantial sensitivity. It also persists when replacing the TATT intrinsic-alignment prescription with the simpler NLA model, suggesting that it is not driven by the specific choice of intrinsic-alignment modelling.

The different reconstructed behaviours of $\mu$ and $\Sigma$ translate into an inferred gravitational slip, $\gamma$, that tends to lie below unity for both the Baseline and Baseline+LSS combinations, with a more pronounced departure in the latter case. Although the statistical significance of these features remains modest, they point to potentially interesting deviations in the gravitational response, particularly in the lensing sector, which merit further investigation with future data.

Many mechanisms capable of realizing stable phantom crossing involve additional degrees of freedom or non-minimal gravitational interactions, and can therefore predict correlated signatures in structure
growth and lensing. Our reconstruction provides a
joint observational framework for investigating
such possibilities without imposing a specific
theory-dependent mapping between the background
and perturbation sectors. Although current data
favour a more pronounced departure in the
background expansion than in linear gravitational
dynamics, the reconstructed gravitational responses leave open the possibility of a common physical origin. We explore this question further through
a combined-mode analysis in our companion
Letter~\cite{CompanionPRL}. Ongoing Stage-IV
LSS observations will provide a more precise test
of possible correlated departures from
GR.


\begin{acknowledgments}
We thank Matteo Martinelli for his close interactions and insightful discussions during the initial stages of this project; we also acknowledge useful conversations with Gen Ye, Hamid Mirpoorian and Marco Raveri.
This work was supported by the European Research Council under the H2020 ERC Consolidator Grant “Gravitational Physics from the Universe Large scales Evolution” (Grant No. 101126217 — GraviPULSE).
This work used the Dutch national e-infrastructure with the support of the
SURF Cooperative using grant no. EINF-18786.
\end{acknowledgments}


\appendix

\begin{figure}[tbp]
    \centering
    \includegraphics[width=\columnwidth]{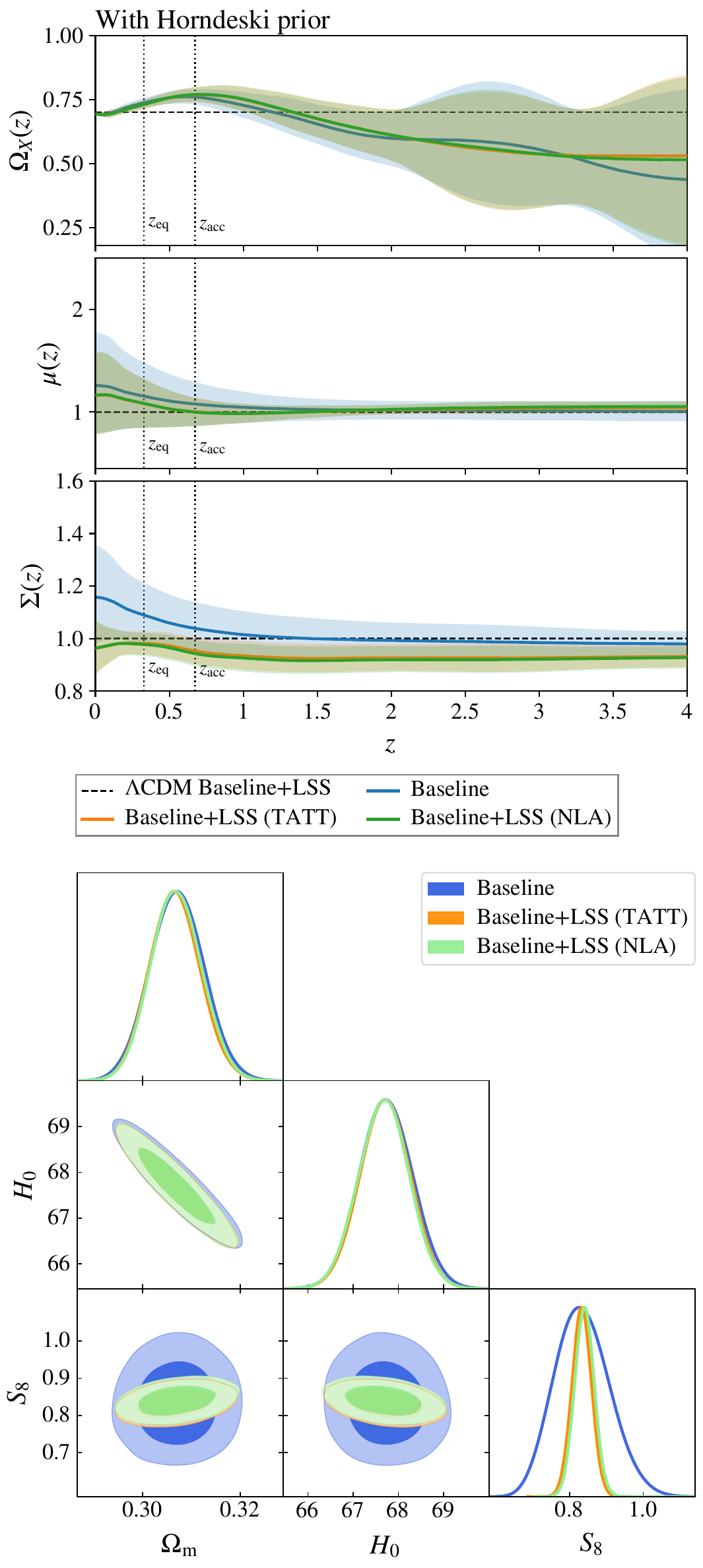}
    \caption{
    Sensitivity of the reconstructed $\Omega_{\rm X}(z)$, $\mu(z)$, and $\Sigma(z)$ to intrinsic-alignment modelling (upper panels).
    Marginalized posterior distributions and $68\%$ and $95\%$ credible contours for $\Omega_{\rm m}$, $H_0$, and $S_8$ (lower panel). Orange and green show Baseline+LSS using TATT and NLA, respectively; blue shows Baseline for reference. All analyses include the Horndeski correlation prior.
    }
    \label{fig:ia_reconstruction}
\end{figure}

\section{Sensitivity to intrinsic-alignment modelling}
\label{app:ia}
Intrinsic alignments (IA) can partially mimic changes in the gravitational response inferred from weak-lensing observables. We assess the sensitivity of our reconstruction to the IA prescription by comparing two widely used models: nonlinear alignment (NLA) and tidal alignment and tidal torquing (TATT)~\cite{DES:2026fyc}. NLA describes the intrinsic-shape field through a linear response to the tidal field, extending the linear-alignment prescription through the use of nonlinear power spectra. In our NLA implementation, we account for the modified lensing response by evaluating gravitational--intrinsic correlations from the matter--Weyl power spectrum, while retaining the standard matter-based expression for intrinsic--intrinsic correlations. TATT extends this description by including quadratic tidal-torquing and density-weighting contributions, allowing greater flexibility in the scale and redshift dependence of IA.

We repeat the prior-assisted Baseline+LSS reconstruction using NLA in place of TATT. The upper panels of Fig.~\ref{fig:ia_reconstruction} compare the functions obtained with TATT and NLA, and include the prior-assisted Baseline case for reference. The two Baseline+LSS analyses show closely overlapping posterior bands for $\Omega_{\rm X}$, $\mu$, and $\Sigma$. In particular, the low-redshift maximum of $\Omega_{\rm X}$ and the tendency towards $\Sigma<1$ persist under the change of IA prescription. The lower panel of Fig.~\ref{fig:ia_reconstruction} shows similarly consistent constraints on $\Omega_{\rm m}$, $H_0$, and $S_8$.
Within the posterior uncertainties, the reconstructed features and parameter constraints are consistent between the two IA prescriptions. We therefore retain the more flexible TATT prescription for the main analysis and use the NLA reconstruction as a robustness check. This agreement does not exclude systematic effects shared by both prescriptions.


\bibliographystyle{apsrev4-2}
\bibliography{references}

\end{document}